# Analysis of Dynamic Linear and Non-linear Memristor Device Models for Emerging Neuromorphic Computing Hardware Design

Nathan R. McDonald, Robinson E. Pino, *Senior Member, IEEE*, Peter J. Rozwood, and Bryant T. Wysocki
*Air Force Research Laboratory, Information Directorate, Rome, NY*

***Abstract*—The value memristor devices offer to the neuromorphic computing hardware design community rests on the ability to provide effective device models that can enable large scale integrated computing architecture application simulations. Therefore, it is imperative to develop practical, functional device models of minimum mathematical complexity for fast, reliable, and accurate computing architecture technology design and simulation. To this end, various device models have been proposed in the literature seeking to characterize the physical electronic and time domain behavioral properties of memristor devices. In this work, we analyze some promising and practical non-quasi-static linear and non-linear memristor device models for neuromorphic circuit design and computing architecture simulation.**

## I. Introduction

THE neuromorphic computing hardware community has been re-energized by the discovery of the physical memristor device by researchers at Hewlett-Packard (HP) Laboratories, in Palo Alto, California, in 2008 [1]. The memristor device, whose name comes from the contraction of "memory resistor," has been characterized as the functional equivalent to the synapse [1]. Leon Chua theorized the existence of the memristor device in 1971 as the fourth basic circuit element [2]. Given the non-volatile nature of the memristor device, applications containing such devices lay within memory and computing applications [1]. As mentioned, the memristor device operates analogously to the biological synapse [1]–[3]; therefore, it represents a step forward in the development of low power and large scale neuromorphic computing hardware and applications.

In order to apply memristor device technology to large scale computing systems, it is important to accurately model and simulate its time domain electronic characteristic behavior. Memristor devices exhibit a strong hysteresis; therefore, based on the current device resistance (or memristance) state or initial conditions, we must be able to accurately predict its future electronic behavior. Several models have been proposed in the literature to describe the non-quasi-static electronic time domain characteristic behavior of memristor devices [4], [6], [7]. In this work, we present a memristor modeling simulation analysis and comparison of published linear and non-linear dynamical memristor device models. We believe that a solid understanding of memristor modeling and simulation methodologies will lead to accelerated design and development of memristor powered technologies such as neuromorphic computing hardware.

R. E. Pino, N. R. McDonald, B. T. Wysocki, and P. J. Rozwood are with the Air Force Research Laboratory, Rome, NY 13441 USA (e-mail: robinson.pino@rl.af.mil, Phone: 315-330-7109, Fax: 315-330-2953).

## II. Memristor Device Models

### A. Linear Boundary Drift Model

The linear memristor device model reported by Hewlett-Packard [1][4] states that the effective transport mechanism in $TiO_2$ based memristor devices is through the drift of vacancies originating within an oxygen deficient $TiO_{2-x}$ layer [4]. The $TiO_2$ based memristor devices' physical quasi-static transport mechanisms have been recently described in some detail by Pickett *et al*. [5]. As the oxygen vacancies drift under an applied external electric field, the stoichiometric $TiO_2$ will become doped with the ionized vacancies. Treating the doped (oxygen vacancy rich regions) and undoped regions of the device as a pair of resistors in series, the memristance corresponding to a given boundary position or state, $w$, relative to the device length or thickness $D$ can be described as follows [4]:

$$M(w) = R_{on}\left(\frac{w}{D}\right) + R_{off}\left(1 - \frac{w}{D}\right), \quad (1)$$

where $R_{on}$ is the resistance of the doped region and $R_{off}$ is the resistance of the undoped region. A schematic representation of the memristor device model is shown in Figure 1.

The drift velocity, $v_D$, at which the doped/undoped boundary interface moves is defined as [6]

$$\frac{\mathrm{d}w}{\mathrm{d}t} = v_D = \frac{\eta\, \mu_D\, R_{on}}{D}\, I(t), \quad (2)$$

where the oxygen vacancies have a characteristic drift mobility, $\mu_D$, under any applied bias voltage. $\eta$ indicates the polarity of the memristor, where $\eta = 1$ or $-1$ for a device whose doped region is expanding or shrinking respectively under a positive voltage bias. For example, the memristor device in Figure 1 has an $\eta = 1$ polarity.

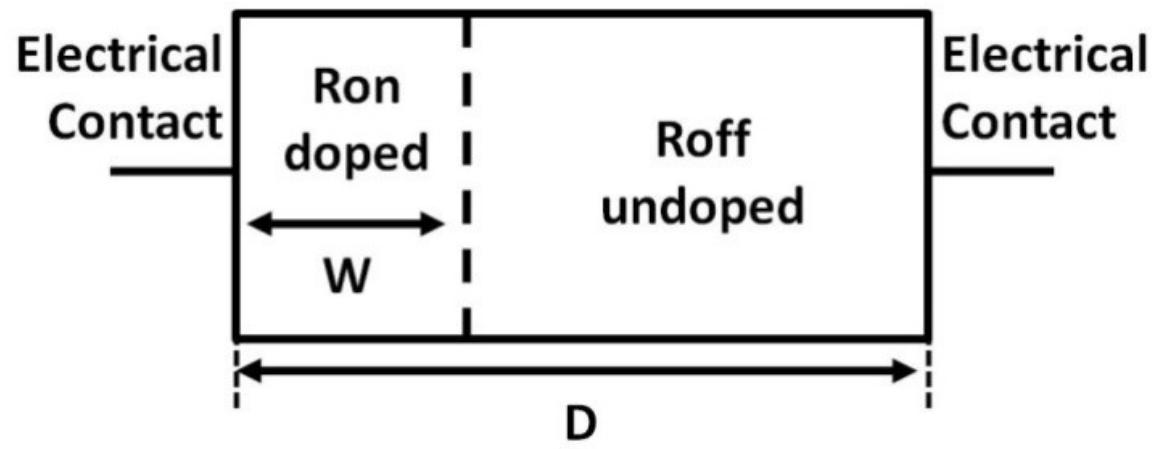

Fig. 1. Schematic representation of the memristor device as two resistors in series.

Integrating both sides of (2) gives the state $w$ as a function of time [6]

$$w(t) = w_0 + \frac{\eta\, \mu_D\, R_{on}}{D} q(t), \tag{3}$$

letting $q(0) = 0$. Substituting (3) into (1), we can solve for the device's memristance, $M$, as a function of charge [6]

$$M(q) = R_0 - \frac{\eta\, \Delta R}{Q_0} q(t), \tag{4}$$

where, after grouping terms, the parameters $R_0$, $Q_0$, and $\Delta R$ are given by [6]

$$R_0 = R_{on}\left(\frac{w_0}{D}\right) + R_{off}\left(1 - \frac{w_0}{D}\right), \tag{5}$$

the initial resistance of the memristor;

$$Q_0 = \frac{D^2}{\mu_D\, R_{on}}, \tag{6}$$

the charge required to alter the state from $w_0$; and

$$\Delta R = R_{off} - R_{on}\,. \tag{7}$$

From Chua's seminal memristance equation [2]

$$\mathrm{d}\varphi = M\, \mathrm{d}q, \tag{8}$$

one may derive essentially Ohm's Law,

$$M\big(q(t)\big) = \frac{\mathrm{d}\varphi/\mathrm{d}t}{\mathrm{d}q/\mathrm{d}t} = \frac{V(t)}{I(t)}\,. \tag{9}$$

Using (4), we can rewrite (9) as [6]

$$V(t) = \left[R_o - \frac{\eta\, \Delta R}{Q_0} q(t)\right]\frac{\mathrm{d}q}{\mathrm{d}t}\,. \tag{10}$$

Then integrating (10) over time, we can solve for the magnetic flux

$$\varphi(t) = R_o q(t) - \frac{\eta\, \Delta R\, q^2(t)}{2Q_0}\,, \tag{11}$$

which, in turn, provides an equation for $q(t)$ via its quadratic solution [6]

$$q(t) = \frac{Q_0\, R_0}{\Delta R}\left(1 - \sqrt{1 - \frac{2\eta\, \Delta R\, \varphi(t)}{Q_0\, R_0^2}}\right), \tag{12}$$

again letting $q(0) = 0$. Substituting (12) into (4), we obtain an equation for memristance explicitly as function of the flux [6]

$$M(q) = R_0\sqrt{1 - \frac{2\eta\, \Delta R\, \varphi(t)}{Q_0\, R_0^2}}. \tag{13}$$

Finally, we can insert (13) into (9) to solve for the current flowing through the memristor device [6]

$$I(t) = \frac{V(t)}{R_0\sqrt{1 - \frac{2\eta\, \Delta R\, \varphi(t)}{Q_0\, R_0^2}}}\,. \tag{14}$$

The linear boundary drift model assumes that the oxygen vacancies are free to traverse the entire length of the move memristor unhindered by the boundary conditions of the device. The utility of this model lies within the ease of usage and closed form solution.

### *B. Non-linear Boundary Drift Models*

The linear boundary drift model reproduces the characteristic time hysteresis behavior of memristor devices; however, the model suffers from oversimplifications of basic electrodynamics. First of all, even small voltages across the nanometer devices will produce a large electric field; thus the ion boundary position will move in a decidedly non-linear fashion. Additionally, $w$ could never reach a zero length because it would indicate that there are physically no oxygen vacancies present in the device, the identified charge transport mechanisms. On the other hand, the entire length of the device could potentially become doped with the oxygen vacancies. Modeling the state change as a mass on a spring, the boundary drift velocity, $v_D$, should be greatest at the center of the device and reduced to essentially zero as $w$ approaches either edge ($w = 0$ and $w = D$). These boundary value restrictions can be implemented by multiplying a windowing function to (2) as shown below [6][7]

$$v_D = \frac{\mathrm{d}w}{\mathrm{d}t} = \frac{\eta\, \mu_D\, R_{on}}{D}\, I(t)\, F(x), \tag{15}$$

where $x = w/D$ is the normalized form of the state variable. The function $F(x)$ should have its highest value at the center of the device ($x = 0.5$) and be zero at the boundaries, $x = 0$ and $x = 1$. Joglekar *et al.* [6] proposed the window function

$$F_p(x) = 1 - (2x - 1)^{2p}, \tag{16}$$

where $p$ is a positive integer. Figure 2 displays a graphical representation of the window function described by (16) for various $p$ solutions ($p$ = 1, 5, and 10). From the figure, we observe that the maximum $F_p(x)$ value occurs at the center of the device and that zero values are obtained at the two boundaries. Also, by varying the $p$ parameter, we can control the rate of change of the function. Lower $p$ values correspond to lower rates of change and vice versa. Inserting (16) into (15), we obtain the modified state change equation

$$\frac{\mathrm{d}w}{\mathrm{d}t} = \frac{\eta\,\mu_D\,R_{on}}{D} I(t)[1 - (2x - 1)^{2p}]\,. \quad (17)$$

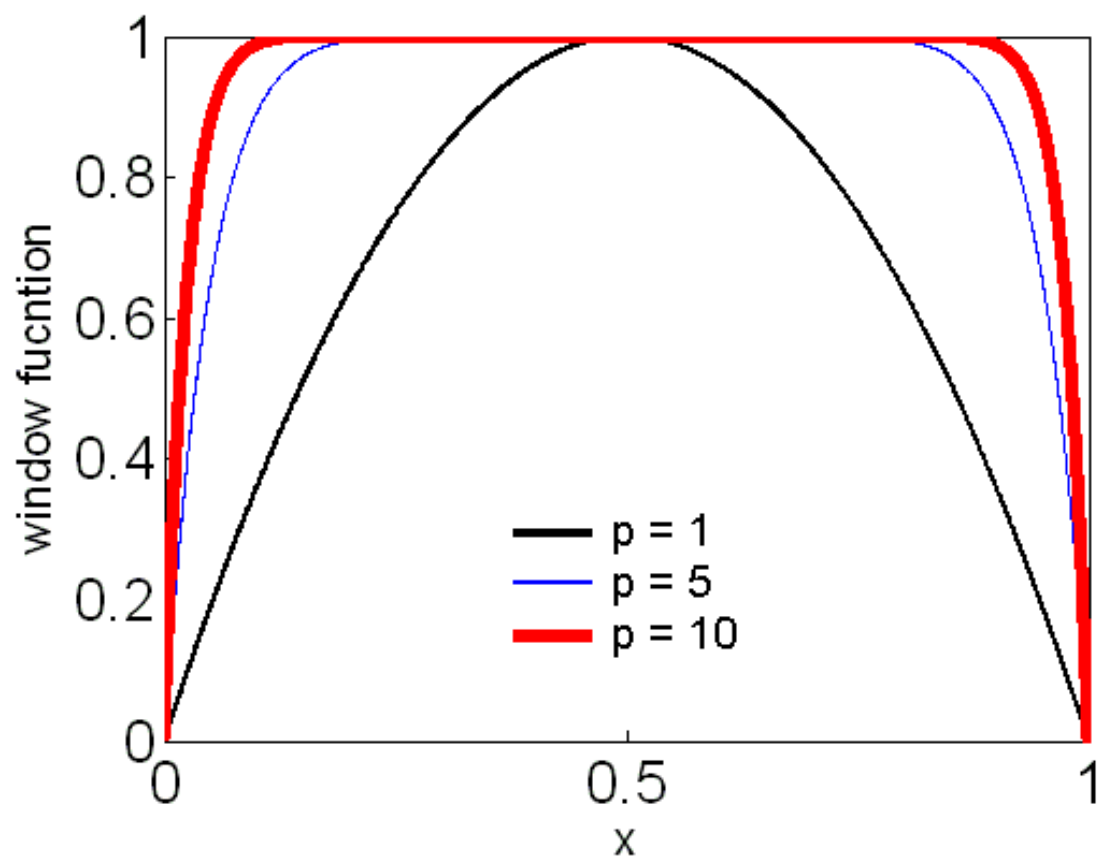

Fig. 2. Plot of non-linear window function proposed by Joglekar *et al.* for $p$ = 1, 5, and 10.

We observe that (17) reduces to the linear boundary drift model described by (2) as $p$ increases [6]. Equation (17) also utilizes the $\eta$ parameter since memristor devices are asymmetric. During modeling and simulation it is important to consistently specify each device's physical orientation.

The non-linear state change model described by (17) is more physically accurate when compared to the linear model; however, the window function makes solving for $w$ as function of time challenging for an arbitrary $p$. Therefore, a time-step numerical solutions approach was employed for simulations. The following formulae were independently derived from the algebraic manipulation of (1), (9), and (17) as shown below

$$M(w(t_i)) = R_{on}\left(\frac{w(t_i)}{D}\right) + R_{off}\left(1 - \frac{w(t_i)}{D}\right), \quad (18)$$

$$I(t_{i+1}) = \frac{V(t_{i+1})}{M(w(t_i))}\,, \quad (19)$$

$$v_D(t_{i+1}) = \frac{\eta\,\mu_D R_{on}}{D} I(t_{i+1})\,F_p\left(\frac{w(t_i)}{D}\right), \quad (20)$$

$$w(t_{i+1}) = v_D(t_{i+1})[t_{i+1} - t_i] + w(t_i), \quad (21)$$

$$q(t_{i+1}) = \Phi(t_{i+1})/\,M(w(t_i)), \quad (22)$$

where $t_i$ in (18) corresponds to the initial time step and $t_{i+1}$ in (19) - (22) the next integral time step.

The order of these time-step equations brought to light another challenge in the implementation of (16), specifically when the doped region covers the entire device length ($x$ = 1). It then follows that $F_p(x = 1) = 0$ for all $p$, (16). Thus, $w$ in (21) does not change since $v_D = 0$, (20). Therefore, $x = 1$ once again for the next time-step during simulation. Then, this loop persists till the end of the simulation without respect to the change in the direction of the current, producing invalid results.

A new window function was proposed by Biolek *et al.* [7]

$$F_p(x) = 1 - \left[x - \mathrm{u}(-I)\right]^{2p}\,, \quad (23)$$

where

$$\mathrm{u}(I) = \begin{cases} 1, & \text{if } I \geq 0 \\ 0, & \text{if } I < 0 \end{cases}\,. \quad (24)$$

This window function is displayed in Figure 3 for various $p$ integer values ($p$ = 1, 5, and 10). The state change is no longer modeled as a mass on a spring; rather, the function is asymmetric in the way it limits changes in $v_D$. For example, when $x$ starts at 0, the function equals 1. Then, as $x$ increases, approaching $D$, the function approaches 0. Once the current reverses direction, the function immediately switches to 1. Then, as $x$ decreases back to 0, the function also decreases to 0. When the current reverses, the cycle begins once again. In order to compute $v_D$, (23) can be substituted into (20) without altering the other four equations. One advantage of Biolek's window function is that it eliminates convergence issues at the device boundaries.

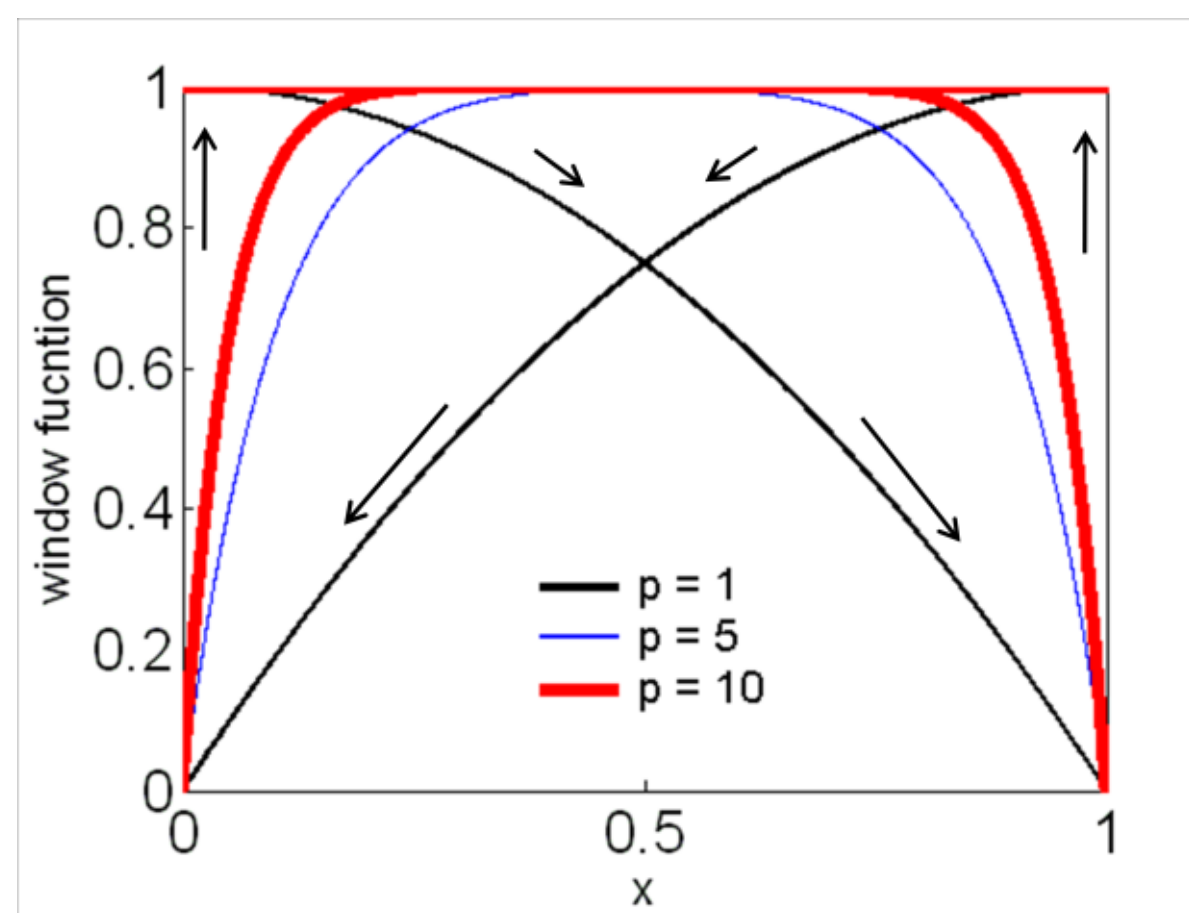

Fig. 3. Plot of non-linear window function proposed by Biolek *et al.* for $p$ = 1, 5, and 10.

## III. Results And Discussion

During model analysis and simulation, all memristor models were simulated in Matlab; and all bias voltage sources were of the form

$$V(t) = v_0 \sin(\omega_0\,t + \theta), \quad (25)$$

where $v_0$ is the voltage amplitude and $\theta$ is an arbitrary phase shift. Typical simulation input parameter values are $v_0$ = 1 – 5 V and $\omega_0$ = 10 – $10^6$ rad/s. We can calculate the flux through the device as the time integral of the voltage across it from (25)

$$\Phi(t) = \left(\frac{v_0}{\omega_0}\right)[\cos\theta - \cos(\omega_0 t + \theta)]. \quad (26)$$

## A. Linear Boundary Drift Model

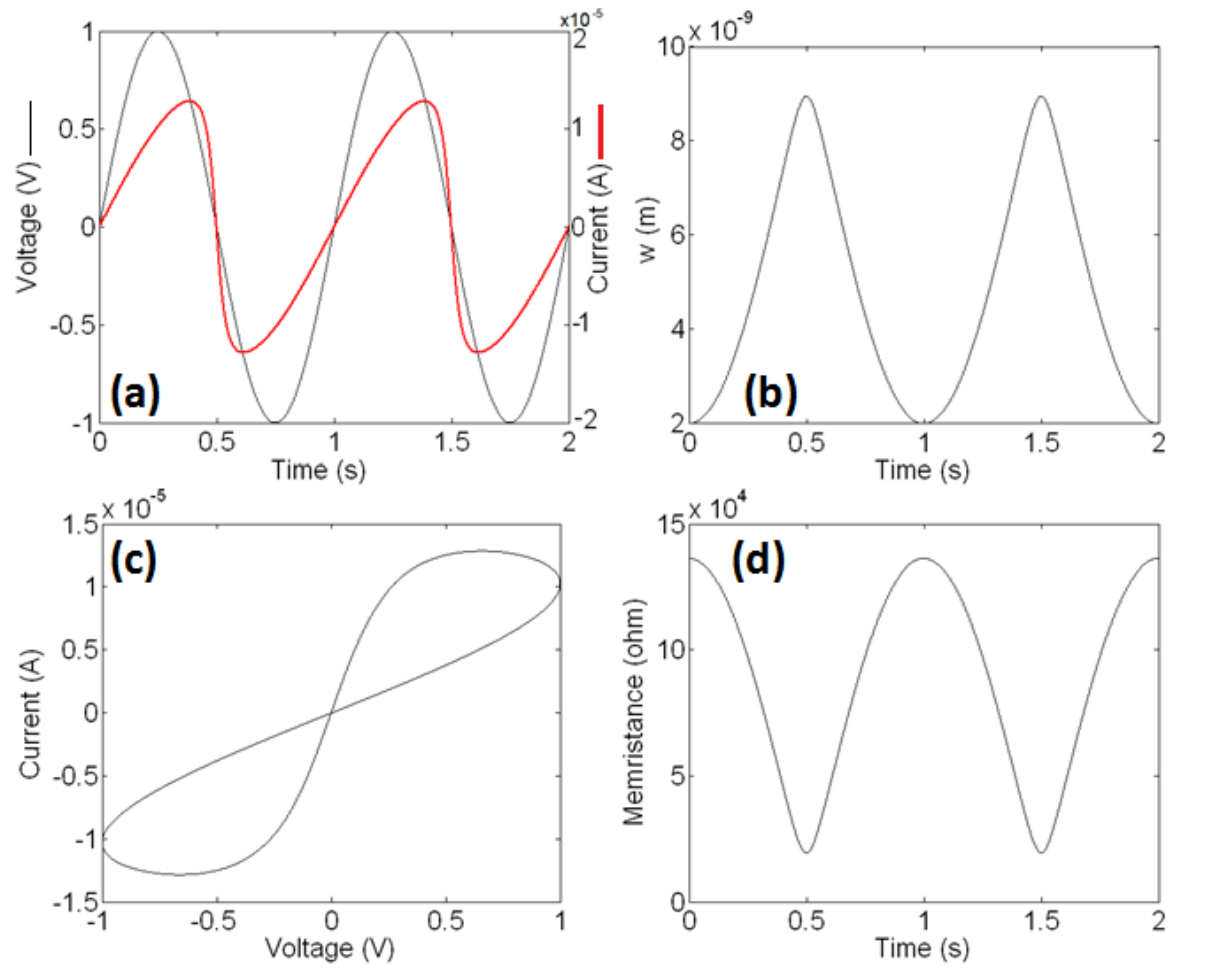

Fig. 4. Plots of $I(t)$ & $V(t)$ (a), $w(t)$ (b), *V-I* hysteresis behavior (c), and $M(t)$ (d) memristor simulation results using the linear boundary drift model, with parameters $\mu_D = 10^{-14}$ $m^2V^{-1}s^{-1}$, $D = 10$ nm, $x_0 = 0.2$, $R_{on} = 1700$ Ω, $r = 100$, $v_0 = 1$ V, $\omega = 2\pi$ rad/s, and $V(t) = \sin(2\pi\ t)$ V.

The physical memristor device is characterized by the parameters $\mu_D$, $w_0$, $D$, $R_{off}$, and $R_{on}$. Adjustments to the dopant mobility parameter directly correlates to changes in the boundary drift velocity as described in (2). A slower (faster) velocity corresponds to smaller (larger) changes in $w$ per cycle, which in turn decreases (increases) the resistance value spectrum available to the memristor. Adjusting $w_0$ also directly alters the effective range of resistance values available to the memristor. In general, a higher $w_0$ produces wider loops in the *I-V* plots. However, neither $\mu_D$ nor $w_0$ can be set to completely arbitrary values; otherwise, imaginary numbers arise in the equations. Overall, the model operates over the widest range of parameter values when the initially doped region is less than half the device length. The maximum viable $\mu_D$ and $w_0$ values are related to the frequency of the voltage source, where a high frequency allows for larger values in both parameters. Long devices, high $D$ values, display less memristive effects than short devices because, as is seen in (4), memristance falls off as an inverse square function.

The $R_{off}$ and $R_{on}$ resistance values can be arbitrarily set in accordance with their definitions. The ratio $r = R_{off} / R_{on}$ should be greater than 10, though ratios of $r = 100 – 2000$ are more commonly used. Increasing $r$ generally reduces the *I-V* curve to a straight line. Additionally, for any given $D$, hysteresis effects are most prominent when $\Delta R >> R_0$ [6]. For the linear state change model, typical parameters were $\mu_D = (10^{-12} – 10^{-14}\ m^2 \cdot V^{-1} \cdot s^{-1})$, $D = (10 – 50$ nm), $x_0 = (0.1 – 0.6)$, $R_{on} = (100 – 1000$ Ω), and $r = (100 - 2000)$.

Figure 4 shows typical simulation results. Figure 4(a) superimposes the input voltage in time (thin line) against the current in time (thick line). From the plot, it is apparent that while the current lags the voltage, both curves have the same period. This shows that the memristor does not store any charge itself but is a totally dissipative circuit element [2]. Figure 4(c) depicts the symmetric, smooth hysteresis loop of an ideal memristor. Figures 4(b) & 4(d) show the variation in width $w$ and memristance over time, respectively. From the figures, we can observe that when $w$ is greatest, memristance is minimum and vice-versa. Both parameters mirror each other.

## B. Non-linear Boundary Drift Models

For modeling and simulation of non-linear memristor models, the optimal time-step values, $\Delta t = t_{i+1} - t_i$ , were determined to be between $10^{-2} – 10^{-4}$ sec. The model simulation results using Joglekar's window function are shown in Figure 5. From the results, we observe that the *I-V* plots, figures 5(a) and (c), exhibit a more pointed signature compared to the linear model results in figures 4(a) and (c). While both $I(t)$ plots have the same period as their respective voltage inputs, figures 5(a) and (c) are sharper because of the usage of the window function. We also noticed, though not shown graphically, that for high $p$ integer values, the non-linear model behaves as its linear counterpart. It is important to notice that the memristance and $w$ plots remain similar for both linear and non-linear models as shown in figures 4 and 5.

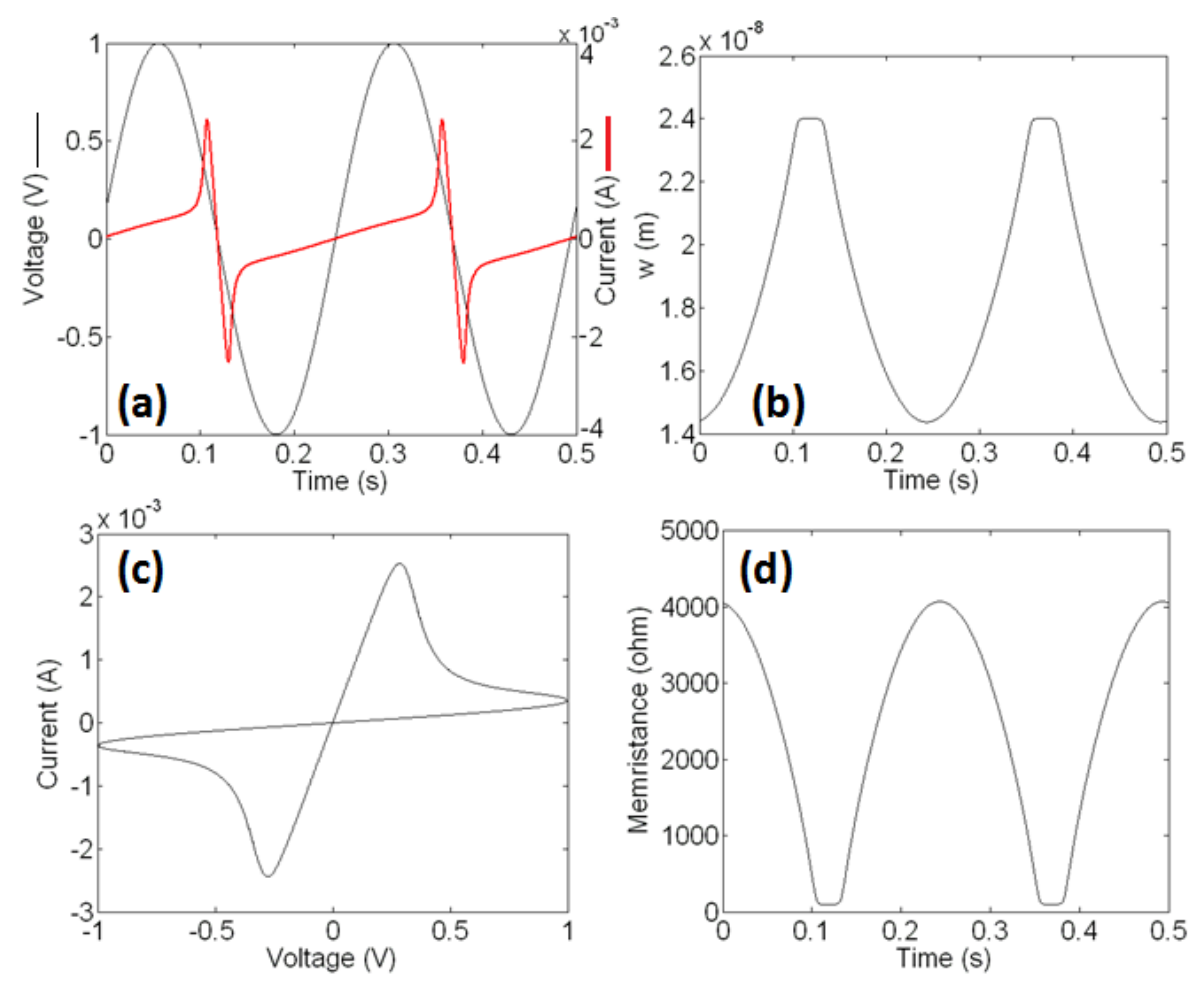

Fig. 5. Plots of $I(t)$ & $V(t)$ (a), $w(t)$ (b), *V-I* hysteresis behavior (c), and $M(t)$ (d) memristor simulation results using non-linear dopant drift model and Joglekar's window function, with parameters $\mu_D = 6.4$ x $10^{-14}$ $m^2V^{-1}s^{-1}$, $D = 24$ nm, $w_0/D = 0.6$, $R_{on} = 100$ Ω, $r = 100$, p = 7, $v_0 = 1$ V, $\omega = 8\pi$ rad/s, $V(t) = \sin(8\pi\ t + 0.16)$ V, $\Phi(t) = \left(\frac{1}{8\pi}\right)$ $[\cos(0.16) – \cos(8\pi\ t + 0.16)]$ Wb, and $\Delta t = 10^{-4}$ sec.

Under certain sets of parameters, the memristor will fluctuate for a few cycles before it settles on a consistent pattern of behavior. However, an appropriate phase shift choice eliminates these initial fluctuations as is shown in the results of Figure 5, where a phase shift of 0.16 rad was employed. The window function also gives the model added robustness in terms of arbitrary parameter range selection. In addition, in terms of parameter selection and adjustment, both linear and non-linear models are similarly affected.

In terms of simulation stability, certain non-linear model simulations cannot be performed for an arbitrary length of time when employing Joglekar's window function. This failure is caused by the convergence issue described in Section II B. To partially remedy this problem for additional simulation time, we could increase $D$ (up to around 50nm, maintaining physical dimensions). However, it is not a comprehensive solution.

In order to circumvent the convergence issues originating from Joglekar's window function, we can employ Biolek's approach described by (23) [7]. Simulation results employing Biolek's window function are displayed in Figure 6. From the figures, we observe that the results preserve the highly non-linear device characteristic behavior. In addition, Biolek's model is unique because it allows for general asymmetric *I-V* device behavior modeling, which is not realizable except in extreme circumstances with the two previous models. This is significant because published physical memristor experimental data [4][5] exhibits asymmetric characteristic behavior.

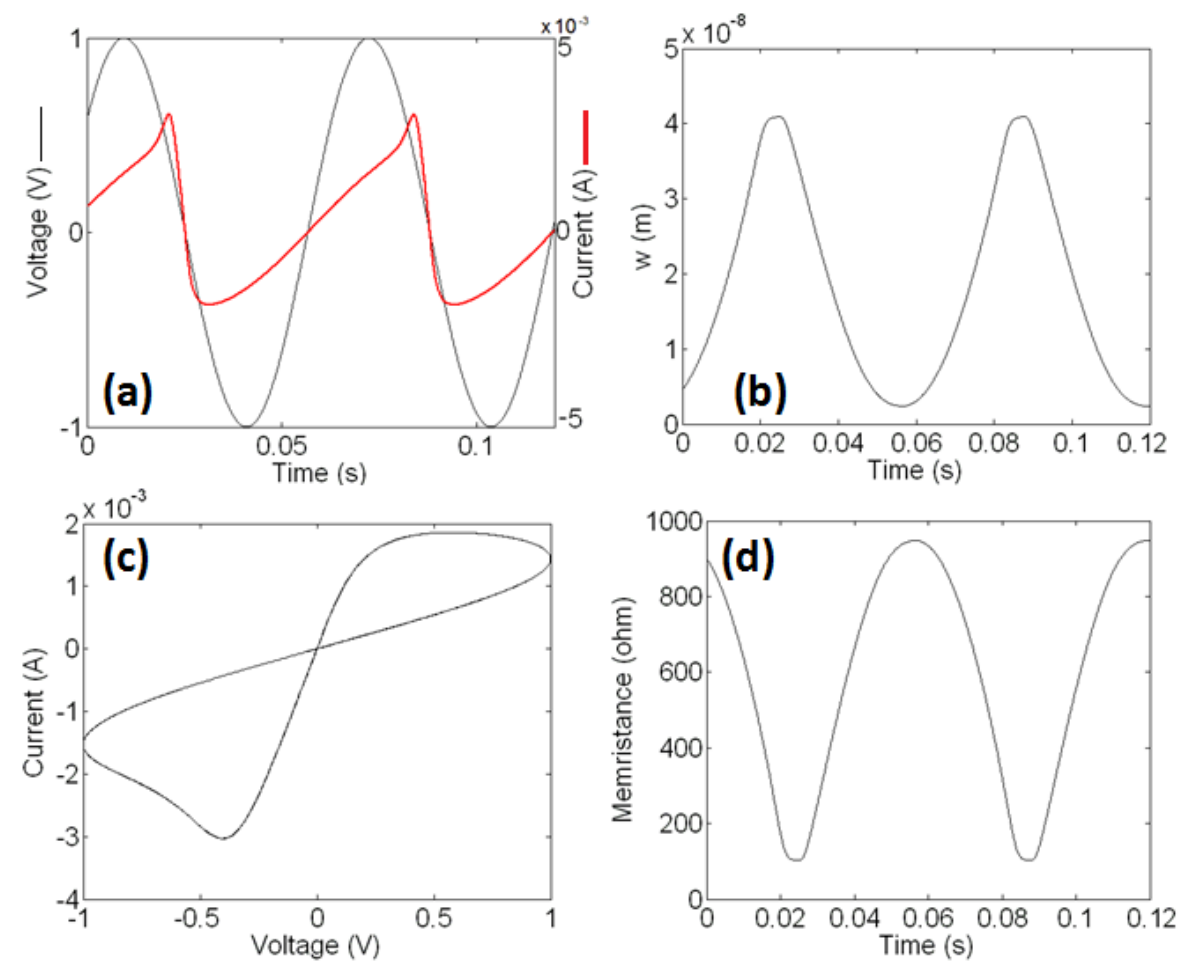

Fig. 6. Plots of $I(t)$ & $V(t)$ (a), $w(t)$ (b), *V-I* hysteresis behavior (c), and $M(t)$ (d) memristor simulation results using non-linear dopant drift model and Biolek's window function, with parameters $\mu_D = 4.4 \times 10^{-13}$ $m^2 \cdot V^{-1} \cdot s^{-1}$, $D = 41$ nm, $w_0/D = 0.11$, $R_{on} = 100\ \Omega$, $r = 10$, p = 7, $v_0 = 1$ V, $\omega = 100$ rad/s, $V(t) = \sin(100\ t + 0.62)$ V, $\Phi(t) = (0.01)[\cos(0.62) - \cos(100\pi\ t + 0.62)]$ Wb, and $\Delta t = 10^{-4}$ sec.

## IV. Conclusion

In this work, we analyzed various published, dynamic linear and non-linear memristor device models. From our study, we observed that the non-linear models offer closer dynamic device characteristic representations when compared to the limited physical published results as opposed to the linear model. The non-linear models, characterized by unique window functions, also provide insight into the dynamics of memristor devices.

Future work will include performing model-to-hardware correlations to physical experimental data when device fabrication is completed. This will provide an opportunity for refining the non-linear memristor models and window functions. Once robust, compact memristor models are in place, circuit level simulations will allow for applications to neuromorphic computing architecture development.

## Acknowledgment

This work was supported by the United States Air Force Research Laboratory, Information Directorate in Rome, NY.

## References

[1] S. Williams, "How We Found the Missing Memristor," *IEEE Spectrum*, vol. 45, no. 12, 2008, pp. 28-35.

[2] L. Chua, "Memristor - The Missing Circuit Element," *IEEE Transactions on Circuits Theory (IEEE)*, vol. 18, no. 5, 1971, pp. 507–519.

[3] L. Chua and S.M. Kang, "Memristive Device and Systems," *Proceedings of IEEE*, vol. 64, no. 2, 1976, pp. 209-223.

[4] D. Strukov, G. Snider, D. Stewart and R. Williams, "The missing memristor found," *Nature,* vol. 453, 2008, pp. 80-83.

[5] M. Pickett, D. Strukov, J. Borghetti, J. Yang, G. Snider, D. Stewart, and R. Williams, "Switching dynamics in titanium dioxide memristive devices," *J. Appl. Phys.*, vol. 106, no. 7, 2009, 074508

[6] Y. Joglekar and S. Wolf, "The elusive memristor: properties of basic electrical circuits," *Eur. J. Phy.*, vol. 30, 2009, pp. 661–675.

[7] Z. Biolek, D. Biolek, and V. Biolková, "Spice Model of Memristor With Nonlinear Dopant Drift," *Radioengineering*, vol. 18, no. 2, 2009, pp. 210-214.